\documentclass[10pt,twoside]{hsqcd}
\usepackage{epsfig}
\usepackage{epsf,amsmath}
\usepackage{graphicx}
 \newcommand{\be}{\begin{eqnarray}}
 \newcommand{\ee}{\end{eqnarray}}
 \newcommand{\eeq}{\end{equation}}
 \newcommand{\ba}{\begin{array}{1}}
 \newcommand{\ea}{\end{array}}
 \newcommand{\bb}{\begin{thebibliography}}
 \newcommand{\eb}{\end{thebibliography}}

\usepackage{color}

\begin{document}

\title{GLUONS IN PROTON AND SOFT $pp$ COLLISIONS AT HIGH ENERGIES }

\author{V.~A.~Bednyakov$^1$, \underline{G.~I.~Lykasov$^2$} and M.~G.~Poghosyan$^{3}$\\ \\
$^{1,2}$ JINR, Dubna, Moscow region, 141980,  Russia, \\
$^3$ Torino University, Torino, Italy\\
$^2$E-mail: lykasov@jinr.ru }

\maketitle

\begin{abstract}
\noindent The hadron inclusive spectra in $pp$ collisions 
at high energies are analyzed within the soft QCD model, namely the quark-gluon 
string model. In addition to the sea quark distribution in the incoming proton 
we consider also the unintegrated gluon distribution that has 
an increasing behaviour when the gluon transverse momentum grows.   
It leads to an increase of the inclusive spectra of hadrons and
their multiplicity in the central rapidity region of $pp$ collision
at LHC energies.

\end{abstract}



\markboth{\large \sl \underline{ G. Lykasov} \& V. Bednyakov, M. Poghosyan
\hspace*{2cm} HSQCD 2010} {\large \sl \hspace*{1cm} GLUONS IN PROTON AND SOFT 
$PP$ COLLISIONS ...} 

\section{Introduction}
As is well known, hard processes involving incoming protons, such as the deep-inelastic lepton-proton
scattering (DIS), are described using the scale-dependent parton distribution functions (PDFs).
Usually such distribution is parametrised as a function of the longitudinal momentum fraction $x$
and the square of the four-momentum transfer $q^2=-Q^2$, integrated over the parton transverse momentum $k_t$.
However, for semi-inclusive processes, such as the inclusive jet production in DIS \cite{Ryskin:2003}, 
electroweak boson production \cite{Ryskin:2003}, etc.,  the parton distributions unintegrated over $k_t$ 
are more appropriate. The theoretical analysis of the unintegrated quark and gluon PDFs is presented recently
in \cite{Ryskin:2010}. According to  \cite{Ryskin:2010}, the gluon distribution function $g(k_t)$ as a function 
of $k_t$ at fixed $Q^2$ has a very interesting behaviour at small $x\leq 0.01$, 
it increases very fast starting 
from almost zero values at $k_t\sim 0$. In some sense, $g(k_t)$ blows up when $k_t$ increases,
then, it falls down at $k_t$ close to 100 Gev$/c$.  In contrast to that the quark distribution $q(k_t)$, as a
function of $k_t$, is almost constant in the whole region of $k_t$ up to $k_t\sim$ 100 GeV$/c$ and smaller 
than $g(k_t)$.
These parametrisations of the PDFs were obtained in \cite{Ryskin:2010} 
within the leading order (LO) and the
next to leading order (NLO) approximations of QCD at $Q^2=10^2$ (GeV$/c$)$^2$ and $Q^2=10^4$(GeV$/c$)$^2$ from known 
(DGLAP-evolved \cite{{DGLAP}}) parton densities determined from global data analysis. 
At small values of $Q^2$ the nonperturbative effects should be considered for properly parametrising the PDFs.
The nonperturbative effects can arise from the complex structure of the QCD vacuum. The instantons are one of the
well studied topological fluctuations of the vacuum gluon fields, see, for example, 
\cite{Shuryak:1998}-\cite{Kochelev:1998}
and references therein. In particular, it is shown \cite{Kochelev:1998} that the inclusion of the instantons
results in the anomalous chromomagnetic quark-gluon interaction (ACQGI) which for the massive quarks gives 
the spin-flip part of it. Within this approach the very fast increase of the unintegrated 
gluon distribution function at $0\le k_t\le 0.5$ Gev$/c$ and $Q^2=1$ Gev$/c$ is also shown.

\section{Inclusive spectra of hadrons in $pp$ collisions}
Let us analyze the hadron production in $pp$ collisions
within the quark-gluon string model (QGSM)  \cite{kaid1} or the dual parton model (DPM) 
\cite{capell1} including the transverse motion of quarks and diquarks 
in colliding protons \cite{LS:1996,LLB:2010}. As is known, the cylinder 
type graphs presented in Fig.1 dominate at high-energy hadronic interactions
\cite{kaid1}. A physical meaning of the graph presented in Fig.1 is the following. The left-hand side 
diagram of Fig.1, the so called one-cylinder graph, corresponds to the 
case when two colorless strings are formed between the quark/diquark ($q/qq$) and the diquark/quark ($qq/q$) 
of the colliding protons, then, at their breaking the quark-antiquark and diquark-antidiquark 
pairs are created and fragmented to hadrons. 
The right hand-side diagram of Fig.1, the so called multi-cylinder graph, corresponds 
to a creation of the same two colorless strings stretched between valence quarks and diquarks
and many strings stretched between sea quarks and antiquarks 
in the different protons.
\begin{figure}[ht]
  {\epsfig{file=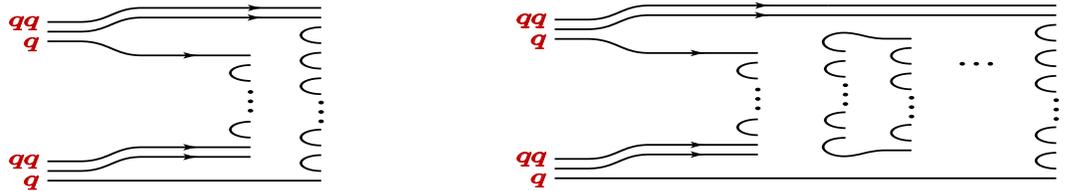,height=2.5cm,width=14.cm  }}
  \caption[Fig.2]{The one-cylinder graph (left) and the multi-cylinder graph (right)
  for the inclusive $p p\rightarrow h X$ process.} 
  \end{figure}

The inclusive spectrum of hadrons in the QGSM is written as follows \cite{LS:1996,
LLB:2010}:
\be
\rho(x,p_t)\equiv E\frac{d\sigma}{d^3{\bf p}}=\sum_{n=1}^\infty \sigma_n(s)\phi_n(x,p_t)~,
\label{def:invsp}
\ee
where $E,{\bf p}$ are the energy and three-momentum of the produced hadron $h$ in the c.m.s. 
of colliding protons, $\sqrt{s}$ is the initial energy; 
$x, p_t$ are the Feynman variable and the transverse momentum of $h$ respectively; $\sigma_n$ is the cross 
section of $n$ cut-Pomeron interaction which is calculated within the ``quasi-eikonal approximation'' 
\cite{Ter-Mart},  
the function $\phi_n(x,p_t)$ has the following form \cite{LS:1996} :
\be
\phi_n(x,p_t)=\int_{x^+}^1dx_1\int_{x_-}^1 dx_2\psi_n(x,p_t;x_1,x_2)~,
\label{def:phin}
\ee
where
\be
\psi_n(x,p_t;x_1,x_2)=F_{qq}^{(n)}(x_+,p_t;x_1)F_{q_v}^{(n)}(x_-,p_t;x_2)/F_{q_v}^{(n)})(0,p_t)+~\\
\nonumber
+ F_{q_v}^{(n)}(x_+,p_t;x_1)F_{qq}^{(n)}(x_-,p_t;x_2)/F_{qq}^{(n)})(0,p_t)+~\\
\nonumber
2(n-1)F_{q_s}^{(n)}(x_+,p_t;x_1)F_{{\bar q}_s}^{(n)}(x_-,p_t;x_2)/F_{q_s}^{(n)})(0,p_t)~.
\label{def:psin}
\ee
and $x_{\pm}=0.5(\sqrt{x^2+x_t^2}\pm x), x_t=2\sqrt{(m_h^2+p_t^2)/s}$,
\be
F_\tau^{(n)}(x_\pm,p_t;x_{1,2})=\int d^2k_t{\tilde f}_\tau^{(n)}(x_\pm,k_t){\tilde G}_{\tau\rightarrow h}
\left(\frac{x_\pm}{x_{1,2}},k_t;p_t)\right)~,
\label{def:Ftaux}
\ee
Here $\tau$ stands for the flavour of quarks and diquarks, ${\tilde f}_\tau^{(n)}(x^\prime,k_t)$
is the quark distribution function that depends on the longitudinal momentum fraction $x^\prime$ and the 
transverse momentum $k_t$. ${\tilde G}_{\tau\rightarrow h}(z,k_t;p_t)=
z{\tilde D}_{\tau\rightarrow h}(z,k_t;p_t)$, where ${\tilde D}_{\tau\rightarrow h}(z,k_t;p_t)$ is the fragmentation
function (FF) of a quark (antiquark) or diquark of flavour $\tau$ into a hadron $h$.  
We assume that both the distribution and the fragmentation functions are factorized
 over the longitudinal and transverse momentum. If their dependence on the transverse momentum has a Gaussian form,
then the function $F_\tau^{(n)}(x_\pm,p_t;x)$ can be written as follows
 \cite{LS:1996}:
\be
F_\tau^{(n)}(x_\pm,p_t;x_{1,2})=f_\tau^{(n)}(x_{1,2})G_{\tau\rightarrow h}(z)
{\tilde I}_n(z,p_t)~,
\label{def:finFtaun}
\ee
where the function ${\tilde I}_n(z,p_t)$ reads \cite{LS:1996}
\be
{\tilde I}_n(z,p_t)=\frac{\gamma_z}{\pi}\exp(-\gamma_z p_t^2),~
\gamma_z=\frac{{\tilde\gamma}}{1+n\rho z^2},~ \rho=\frac{{\tilde\gamma}}{\gamma}~.
\label{def:gammaz}
\ee
Note that at $x=0$ the function $F_\tau^{(n)}$ does not depend on $n$. In our calculations we 
have parametrized the quark distribution and FF by an exponent. In this case the function
${\tilde I}_n$ has a more complicated form. However, in this case $F_\tau^{(n)}$ does not depend on 
$n$ too.  

$\bullet~${\bf Gluon distribution in proton }.\\
As is mentioned in the Introduction, the unintegrated gluon distribution in the proton at small values of
$x$, as a function of $k_t$, increases very fast when $k_t$ increases and then slowly decreases, according
to the instanton vacuum approach for the massive quarks \cite{Kochelev:1998} at low values of $Q^2\sim$ 1 GeV$/c$. 
The similar behaviour for $g(k_t)$ was obtained within the NLO QCD calculations at large $Q^2=10^2$GeV$/c$
and $Q^2=10^4$GeV$/c$ for the massless quarks \cite{Ryskin:2010}.
This allows us to assume that at $Q^2=0$ and  $x=0$ the gluon distribution in proton, as a function of $k_t$,
has a similar behaviour.
\be
g(k_t)\sim k_t^{a_g}\exp(-b_g k_t)~,
\label{def:gkta}
\ee
where the parameters $a_g>0$ and $b_g>0$.

$\bullet~${\bf Hadron production in central rapidity region}.\\
According to the Abramovskiy-Gribov-Kancheli cutting rules (AGK) \cite{AGK}, 
at mid-rapiditiy only Mueller-Kancheli type diagrams contribute to the inclusive spectrum
of hadrons. In our approach the function $F_\tau^{(n)}$ is parametrized on such a way, 
that in the central region ($y=0$), when $x\simeq 0$ and $z\simeq 0$,
it becomes proportional to $n$ and satisfies to the AGK cancellation. Thus,
\be
\rho_q(x=0,p_t)=\phi_q(0,p_t)\sum_{n=1}^\infty n \sigma_n(s) =
gs^{\Delta}\phi_q(0,p_t),
\label{def:invspq}
\ee  
where $\phi_q(x=0,p_t)$, depends only on $p_t$.
Considering the gluons from incoming protons, which may split into  $q\bar{q}$ pairs, we get an
additional contribution to the spectrum:
\be
\rho_g(x=0,p_t)=\phi_g(0,p_t)\sum_{n=2}^\infty 
(n-1)\sigma_n(s)\equiv \phi_g(0,p_t)(gs^{\Delta}-\sigma_{nd})~,
\label{def:invspg}
\ee
Where $\Delta=0.12$, g$=$21 mb and $\sigma_{nd}$ is the cross section of any number cut-Pomeron
exchange.
The quantities
\be
\sum_{n=1}^\infty n\sigma_n(s)=gs^{\Delta}~;~
\sum_{n=1}^\infty\sigma_n(s)=\sigma_{nd}
\label{def:gsignd}
\ee
were calculated in \cite{Ter-Mart}  within the ``quasi-eikonal'' approximation \cite{Ter-Mart}.
Assuming that one of the cut-Pomerans is always scratched between valence quarks and diquarks
which are not coming from the splitting of gluons, in Eq.~(9) we excluded unity from $n$.
Finally, we can present the inclusive spectrum at $x\simeq 0$ in the following form:
\be
\rho(p_t)=\rho_q(x=0,p_t)+\rho_g(x=0,p_t)=
gs^{\Delta} \phi_{q}(0, p_t)+\left(gs^{\Delta}- \sigma_{nd}\right)
\phi_g(0, p_t)~,
\label{def:rhoagk}
\ee    
We fix these contributions from data on the charged particles $p_t$ distribution, 
parametrising them as follows:
\be
\phi_q(0,p_t)=A_q\exp(-b_q p_t)~,~ \phi_g(0,p_t)=A_g\sqrt{p_t}\exp(-b_g p_t).
\label{def:phiq}
\ee
The parameters are fixed  from a fit to data on $p_t$ distribution of charged particles at $y=0$:
  $A_q=4.78\pm 0.16; b_q=7.24\pm 0.11$ and $A_g=1.42\pm 0.05; b_g=3.46\pm 0.02$.

\section{Results and discussion}
In Fig.2 we illustrate how our approach works for the description of the ISR data on inclusive 
spectra of charged pions and kaons. The solid lines correspond to our calculations without the gluon
contribution $\rho_g$. As one can see, we have obtained a quite good description of the spectra
as functions of $p_t$, up to $p_t=1.4$ GeV$/c$ at different values of $x$. Though, in Fig. 2 we 
present results for $\sqrt{s}=53$ GeV, but the description of data at other ISR energies 
($\sqrt{s}=23.3, 30.6, 44.6, 53$ GeV) is similarly good. 
The result of the fit to data on the charged hadron inclusive spectra
is presented in Fig.3. The long dashed curve corresponds to the quark contribution $\rho_q(x=0, p_t)$ given by 
Eq.(\ref{def:invspq}), whereas the short dashed line is the gluon contribution $\rho_g(x=0, p_t)$ (Eq.(\ref{def:invspg}))
to the invariant yield $d^3N/dyd^2p_t$; the solid curve corresponds to the sum of both contributions, see
 Eq.(\ref{def:phiq}).
One can see that the conventional quark contribution $\Phi^q(y=0,p_t)$, see Eq.(\ref{def:phiq}),
is able to describe the data up to $\leq$1 GeV$/c$, whereas the inclusion of the gluon contribution allows 
us to extend the range of good description up to 2 GeV$/c$. At larger values of $p_t$ the contribution of hard
processes is not negligible and one has to take them into account based on the perturbative QCD.

\section{Conclusion}
Our study has shown that the soft QCD or the QGSM is able to describe the experimental data on inclusive spectra of light
hadrons like pions and kaons produced in $pp$ collisions at not large values of the transverse momenta $p_t\leq 1 GeV/c$
rather satisfactorily. The inclusion of the unintegrated distributions of gluons in the proton due to instanton vacuum
excited in the strong $pp$ interaction allows us to extend the satisfactory description of the experimental data on
the inclusive spectra of charged hadrons at $y=0$ up to $p_t\sim 2 GeV/c$. At higher values of
the transverse momentum the hard parton interactions should be considered for describing the experimental data.   
\begin{figure}[ht]
{\epsfig{file=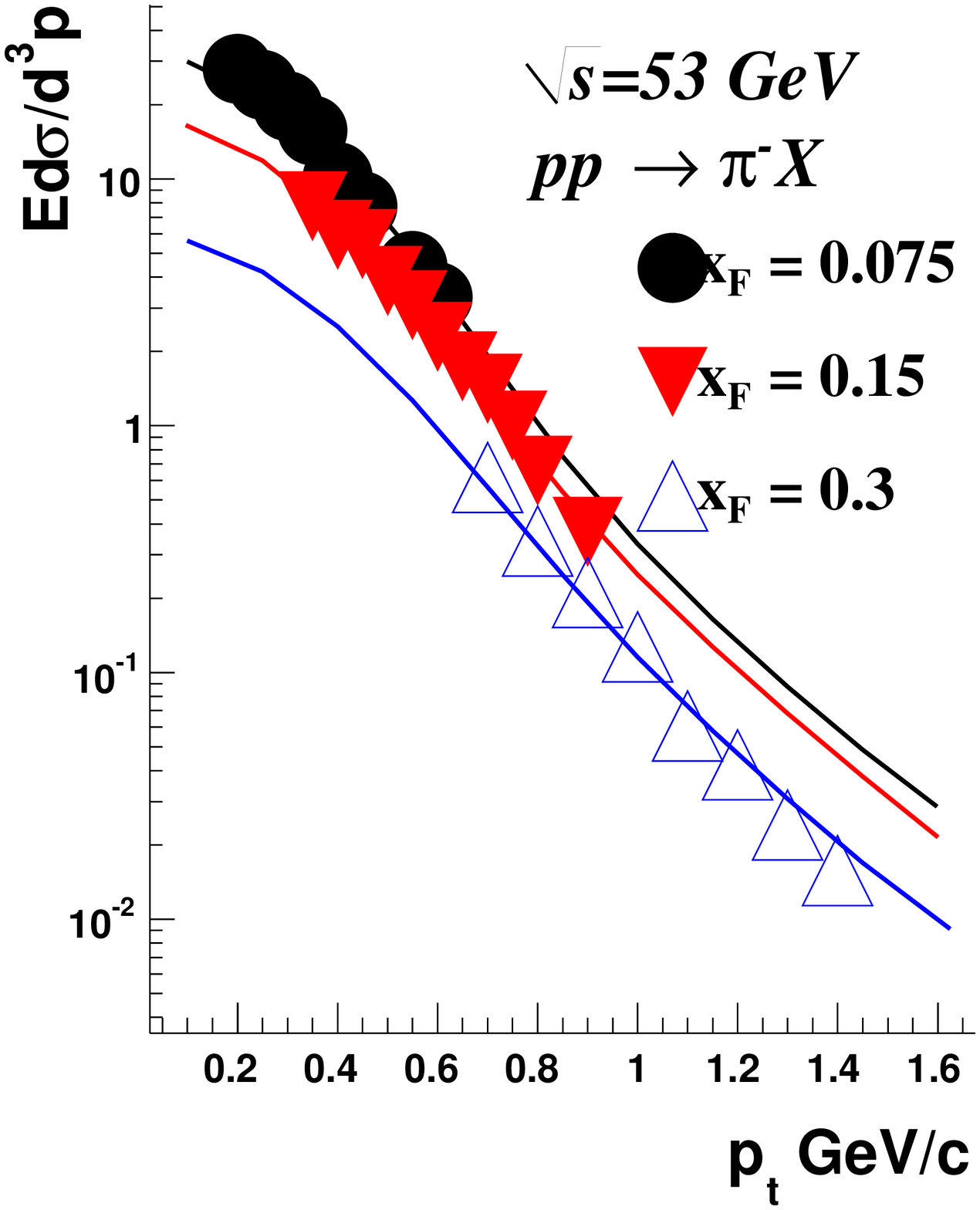,height=6.5cm,width=6.5cm}}
{\epsfig{file=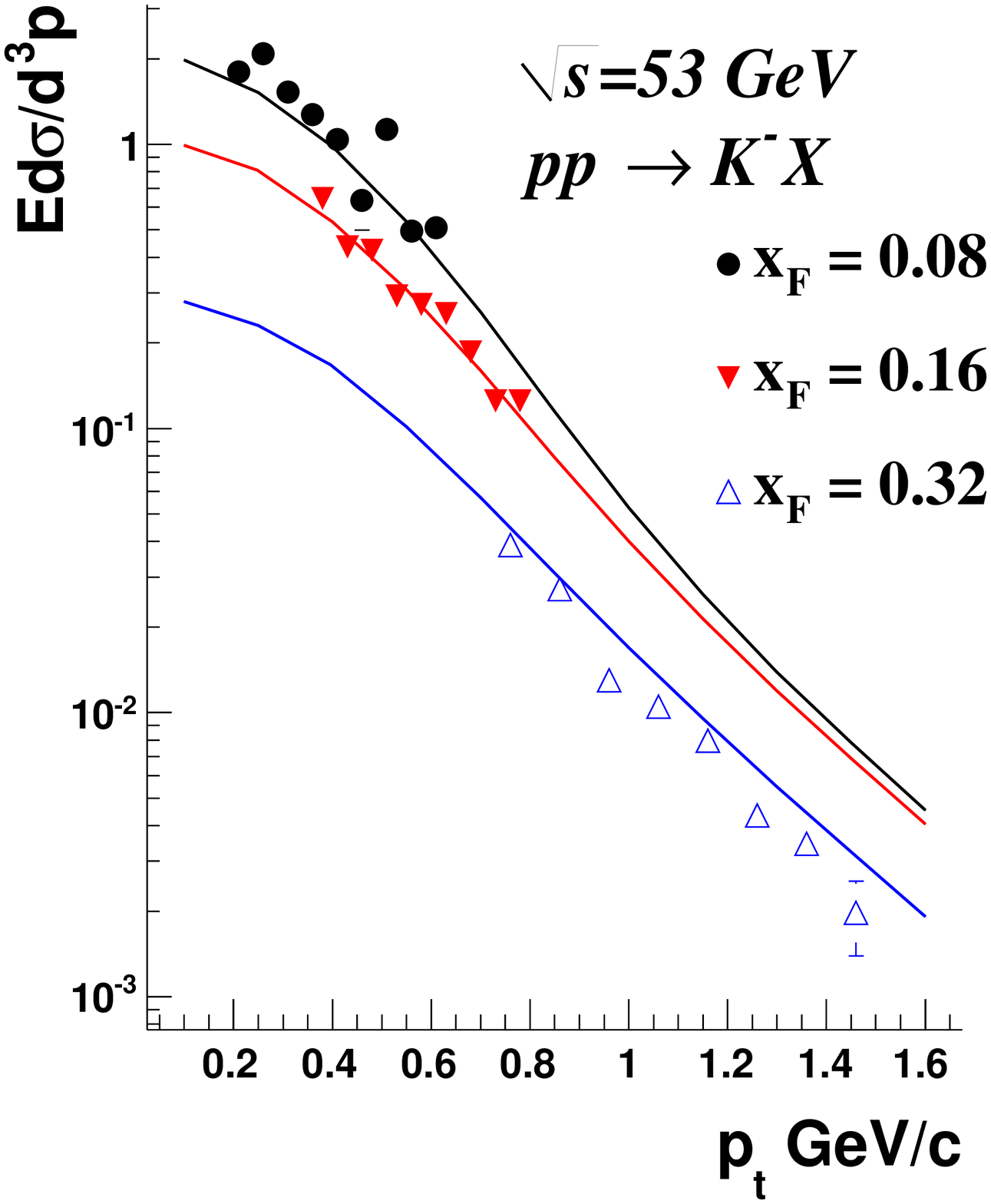,height=6.5cm,width=6.5cm}}
\caption[Fig.2]{The inclusive spectra of $\pi^{-}$ (left) and $K^{-}$ (right) 
mesons in $pp$ collision  $Ed\sigma/d^3p$ [mb GeV$^{-2}$c$^3$] at $\sqrt{s}=$53 GeV compared with the ISR 
data \cite{ISR}. The solid lines correspond to our calculations whithout the gluon
contribution $\rho_g$}.
\end{figure}

\begin{figure}[ht]
{\epsfig{file=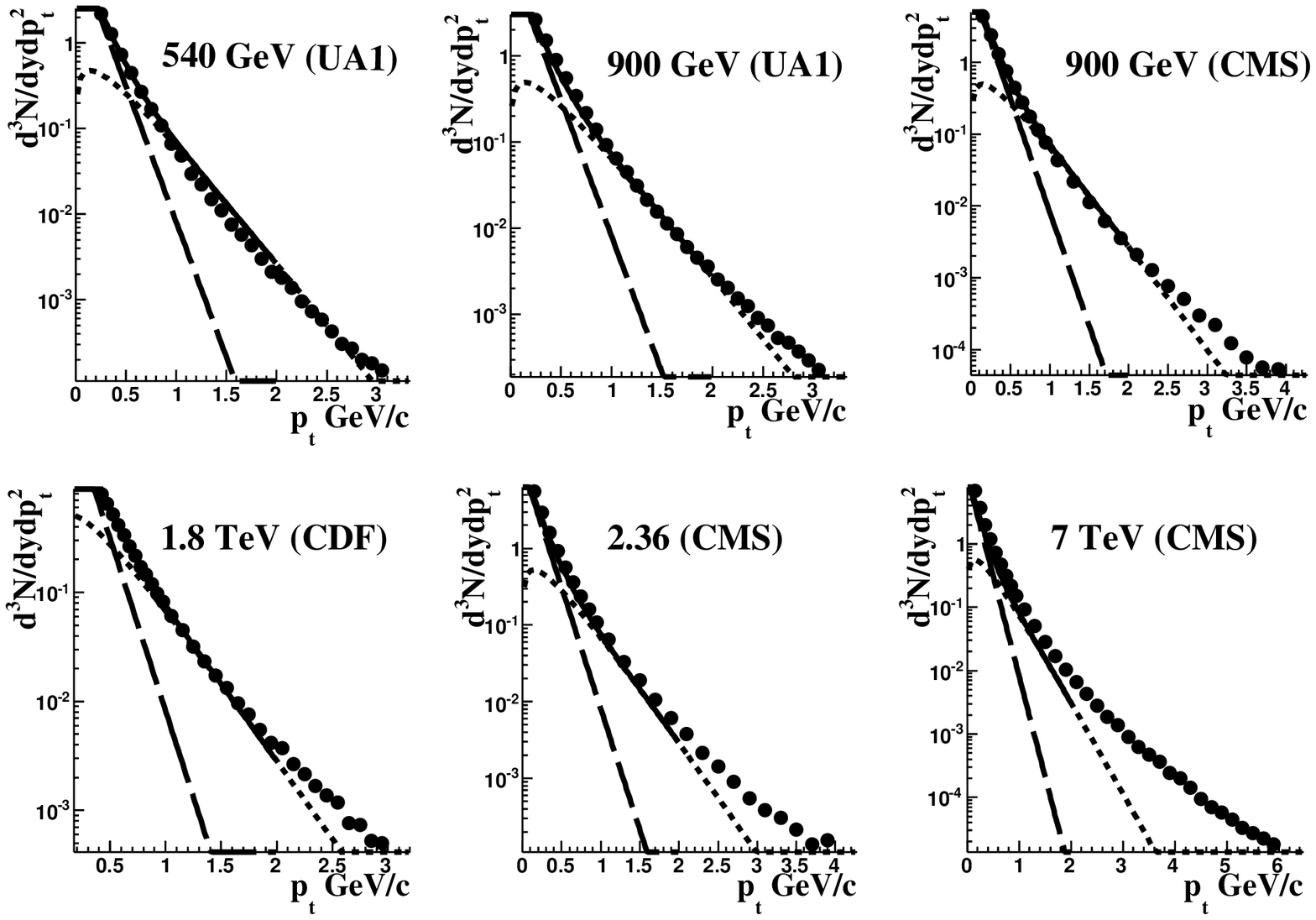,height=11cm,width=14cm  }}
\caption[Fig.3]{
The charged particles yield in the central rapidity region ($y=0$) as 
a function of $p_t$ at $\sqrt{s}$=540 and 900 GeV (top) and 
$\sqrt{s}$=1.8, 2.36 and 7 TeV (bottom) compared with UA1, CDF and CMS data 
 \cite{UA1,CDF,CMS}.
The long dashed curve corresponds to the quark contribution $\rho_q(x=0, p_t)$ given by 
Eq.(\ref{def:invspq}), whereas the short dashed line is the gluon contribution $\rho_g(x=0, p_t)$ (Eq.(\ref{def:invspg}))
to the inclusive yield $d^3N/dyd^2p_t$; the solid curve corresponds to the total calculation including both 
these contributions, see
 Eq.(\ref{def:phiq}).}
\end{figure}
{\bf Acknowledgements}\\
The authors are very grateful to A.Bakulev, A.E.Dorokhov, \frame{A.B.Kaidalov}, O.Kancheli, V.Kim, 
N.Kochelev, Yu.Kulchitsky, E.Kuraev, L.Lipatov, S.Mikhailov, C.Merino and M.Ryskin for very 
useful discussions and comments. 
This work was supported in part by the Russian Foundation for Basic Research 
grant N: 08-02-01003.

\end{document}